\documentstyle[12pt,epsf]{article}
\newcommand{\beq}{\begin{equation}}
\newcommand{\eeq}{\end{equation}}
\newcommand{\bea}{\begin{eqnarray}}
\newcommand{\eea}{\end{eqnarray}}

\newcommand{\CG}{{\cal G}}

\newcommand{\CD}{{\cal D}}

\newcommand{\gstr}{\gamma_{str}}
\newcommand{\pa}{\partial}

\begin{document}
\topmargin 0pt
\oddsidemargin 5mm
\headheight 0pt
\topskip 0mm

\addtolength{\baselineskip}{0.20\baselineskip}

\pagestyle{empty}

\begin{flushright}
OUTP-98-28P\\
30th April 1998\\
hep-th/9805002
\end{flushright}

\begin{center}

\vspace{18pt}
{\Large \bf A Note on the $c=1$ Barrier in Liouville Theory}

\vspace{2 truecm}

{\sc Jo\~ao D. Correia\footnote{e-mail: j.correia1@physics.ox.ac.uk}}

\vspace{1 truecm}
{\em Department of Physics, University of Oxford \\
Theoretical Physics,\\
1 Keble Road,\\
 Oxford OX1 3NP, UK\\}

\vspace{3 truecm}

\end{center}

\noindent
{\bf Abstract.} The instability of Liouville theory coupled to $c>1$
matter fields is shown to persist even when the  ``spikes'' which
represent highly singular geometries are allowed to interact in a
natural way.

\vfill
\begin{flushleft}
\end{flushleft}
\newpage
\setcounter{page}{1}
\pagestyle{plain}

\section{Introduction}

Recent years have seen a great deal of work on non-critical bosonic
string theories, especially using discrete methods
\cite{bowickrev}. Most of the efforts have been directed at
understanding the nature the $c=1$ barrier, where the critical
exponents predicted by the KPZ formulae \cite{usual} become
meaningless. 

As has been pointed out by Bowick, among others, it is unusual that
whatever progress that has been made in understanding this barrier has
been obtained via discrete methods, such as discretized random
surfaces, rather than by considering continuum models. The recent
results of David \cite{scenario} have their origin in the study of
matrix models, and computation of physical quantities in the $c>1$
regime take as their starting point the branched polymer
(BP) ensemble, a collection of simplicial trees which approximate the
highly singular geometry expected to dominate such regions.

Work on the continuum version of the BP transition goes back to the
heuristic arguments of Cates \cite{cates} and the more elaborate
calculations of Krzwicki \cite{k}, in which they considered a BP-like
configuration (a ``spike''), and found that such configurations were
favoured as soon as the central charge $c$ exceeds unity. This spikes
were non-interacting, and Krzwicki in particular noted that it would
be interesting to consider interacting spikes with a view to
determining whether or not the interaction changes the transition.

The purpose of this paper is to consider one such case. We start by
reviewing the relevant arguments from \cite{k} which lead to the
computation of the spike free energy. We then introduce a natural
spike interaction and investigate its properties, as well as computing
the free energy of the interacting spikes in the ''spike gas'' regime.
We end by speculating on the strongly-coupled spike ensemble and its
effect on the BP transition.

\section{Spikes in Liouville Theory}

In non-critical bosonic string theory (or alternatively, in
two-dimensional quantum gravity coupled to matter fields), we
integrate over all 2D geometries and over all matter field
configurations to obtain quantities of interest, such as correlation
functions. Among the more interesting types of matter field we can
couple to geometry are the Conformal Field Theories (CFTs), which are
characterized by the central charge $c$. Once we
integrate out the matter fields, we are left with the Liouville action
\beq S_{L} = \frac{26-c}{96 \pi} \int d^2 \xi \left(\phi \pa^2 \phi +
k e^{\phi} \right) \label{sliouv} \eeq
where $\xi=(\xi_1,\xi_2)$ are the coordinates describing the
manifolds, $\phi(\xi)$ is the Liouville field describing the metric
via $g_{\mu \nu} = e^{\phi} \delta_{\mu \nu}$, and $k$ is the
cosmological constant. From the work carried out in \cite{usual}, we
know that the large-area behaviour of the generating functional 
\beq Z(A) = e^{k_c A} A^{\gstr(h)-3}(1 + \ldots) \label{za} \eeq
is controled by $c$ via the famous KPZ relation
\beq \gstr(h) = 2 - \frac{1-h}{12} \left\{ 25 - c
+\sqrt{(1-c)(25-c)}\right\} \label{kpz1} \eeq
Clearly, formula (\ref{kpz1}) 
makes no sense for
$c>1$, and this breakdown has been attributed to the emergence of BP
configurations which dominate that region. The continuum analogue of
such configurations is Cates' spike
\beq \phi_0 = -\frac{\mu}{2} \log \left\{ (\xi - \xi_0)^2 + \alpha^2
\right\} \label{spike} \eeq
where $\mu>2$ and $\alpha<<1$. The area of such a spike goes like 
\beq A = \int d^2 \xi e^{\phi_0} \sim \alpha^{2-\mu} \label{area} \eeq
up to regular terms in the $\alpha \rightarrow 0$ limit. The question
posed (and answered) by Cates was whether or not such spikes would be
favoured in the $c>1$ region; the answer was affirmative
\cite{cates}. 
The rigorous arguments which confirmed the answer are due to Krzwicki
\cite{k}, and we now give a brief overview of his reasoning.
The action corresponding to the spike (\ref{spike}) can be easily
computed and is, up to regular terms in $\alpha$
\beq S_{\phi_0} = \frac{26-c}{96}\left\{ \mu^3 \log (1/\alpha)
\right\} \label{spikeaction} \eeq
However, we can get further contributions to the free energy from the
functional integration over spike configurations. As shown in
\cite{k}, by considering a suitably regularized Laplacian operator on
a manifold M, the functional integration reduces to a product of
integrations over the centers of the spikes and of integrations of
small fluctuations about the spike configuration, which we write
symbolically as $ \CD \chi \ \prod_i   d^2 \xi_i$.
The expression for $\CD \chi$ can be computed, and it renormalizes the
free energy which becomes 
\beq  S_{\phi_0} = \mu \log (1/\alpha) \left\{ \frac{(25-c)\mu^2}{96}
- 1 -\frac{2(2-\mu)}{\mu} \right\} \label{free} \eeq
Notice the factor of $25-c$, instead of the $26-c$ of
(\ref{spikeaction}), as well as the extra term.

By setting $\mu=2 + \eta$ it is easy to see that (\ref{free}) becomes
large negative if $c>1$ for $\eta$ small; since $\eta$ is arbitrary
(because we expect all spikes to be present in the functional
integration), we conclude that the formation of spikes with $\mu$
close to $2$ is favoured as soon as $c$ becomes greater than unity.

Due to the decoupling of the functional integration, the results above
extend trivially to a gas of $N$ non interacting spikes, which can be
represented by 
\beq \phi = - \sum_{i=1}^{N} \frac{\mu}{2} \log \left\{ (\xi_i -
\xi_{0_i})^2 + \alpha^2\right\} \label{spikegas} \eeq
One could, however, consider interacting spikes, and if we make the
restriction that such interaction is only dependent on the distance
between the centers of the the spikes and not on the shape of the spike
configuration, then the free energy will be changed by an amount
which can be computed from
\beq Z_{int} = \int \prod_{i} d^2 \xi_i e^{-S_{int}
(\xi_{0_1},\ldots,\xi_{0_N})} \label{zint} \eeq

\section{Interacting Spikes}

We consider the case of spikes interacting via a Coulomb-like
potential in two dimensions
\beq S_{int} = \prod_{i \neq j} \CG |\xi_{0_i}-\xi_{0_j}|
\label{inter} \eeq
The coupling constant $\CG$ is chosen to be the product of the areas
of the spikes, which gives an indication of the ``mass'' of each
configuration
\beq \CG = 4 \alpha^{2(2-\mu)}, \label{defg} \eeq
the factor of $4$ being inserted for later convenience \footnote{
While this choice of interaction might appear somewhat restrictive, 
the same qualitative results
would be obtained by considering any analytic function of $\alpha$ and $\eta$
which obeys $f=1$ for $\eta\rightarrow 0^+$, $\alpha>0$.}.

With this choice of interaction, we are left with the task of computing
integrals of the type
\beq Z_{int} (\CG) = \int \prod_l d^2 \xi_{0_l} \prod_{i \neq j}
|\xi_{0_i}-\xi_{0_j}|^{-\CG} \label{zint2} \eeq
which must be regularized. One possibility is to consider the
area of the integration region to be large but limited, another is to
prolong the definition of the integrals (\ref{zint2}) to the complex
plane \cite{complex} and use the conformal symmetry which they then possess to
evaluate them, by fixing three of the spikes at coordinates $0$, $1$
and $\infty$. With this proviso, and introducing $\rho=-\CG/4$ we find
\cite{abdalla} 
\beq Z_{int}(\rho) \simeq (\Delta(1-\rho))^N
\prod_{j=1}^{N} \Delta(j \rho) \prod_{l=0}^{N-1}
(\Delta(1+(l+1/2)\rho))^2 \Delta(-1-(N+l)\rho) \label{bingo} \eeq
where 
\beq \Delta(x)=\frac{\Gamma(x)}{\Gamma(1-x)} = \frac{1}{\pi}
(\Gamma(x))^2 \sin(\pi x) \label{delta} \eeq
and $N$ is the number of free spikes, i.e. those not fixed at special
positions; the total number of spikes is $N+3$. 

From (\ref{delta}) it is easy to classify the zeros and singularities
of (\ref{bingo});
this structure and its
origins have been discussed at length in \cite{abdalla}. The main
point to retain is that the singularity at $\rho=-2$ signals a
transition from an unclumped phase ($\rho>-2$) where the spikes are
distant from each other, to a clumped phase ($\rho<-2$), where they
are very close.
Using (\ref{bingo}) and the properties of the Gamma function, it is
now easy to write down the free energy for some cases of interest:
small coupling ($\rho \sim 0$), and close to the clumping transition
($\rho=-2+\epsilon$ with $\epsilon$ small but positive.) 

We begin by considering $N=1$, the first case for which (\ref{bingo})
is applicable, and we find
\beq Z_{int}=
\left[\frac{\Gamma(1+\rho/2)}{\Gamma(-\rho/2)}\right]^2
\frac{\Gamma(-1-\rho)}{\Gamma(2+\rho)} \label{1} \eeq
which, for small $\rho$ has the expansion
\beq Z_{int} = \frac{\rho}{4} + O(\rho^2) \label{srho1} \eeq
while for $\rho=-2+\epsilon$ it behaves as
\beq Z_{int} =  4 \epsilon^{-1} + O(\epsilon^0) \label{rho-21} \eeq
In both cases the free energy, written in terms of $\alpha$ and $\mu$
reads, retaining only the leading terms,
\beq  F = \mu \log (1/\alpha) \left\{ \frac{(25-c)\mu^2}{96}
- 1 -\frac{4(2-\mu)}{\mu} \right\} \label{free4} \eeq
which should be compared with (\ref{free}); we see that the only
change comes in the last term. This change however does not alter the
previous conclusion, and (\ref{free4}) becomes large and negative for
$c>1$ and small $\eta$. This was to be expected, since at small
coupling the results are not expected to differ dramatically from the
no-coupling regime, as the spikes are widely separated.

The following case of interest is when we allow $N$ to be large. Using
Stirling's formula and the formula for the behaviour of the Gamma
function close to s singularity we find, for small $\rho$
\beq Z_{int} = \frac{2\, \Gamma(N+1/2) 4^{-N}}{\sqrt{\pi}
(\Gamma(N+1))} \rho^N \label{srho2} \eeq
leading to a free energy
\beq F=-N \log(\rho) - N \log(4) +O(\log(N)) \label{freeN1} \eeq
The $O(\log(N))$ order terms are of no consequence to the overall
behaviour of the free energy in the large $N$ limit, the $\log(4)$
term is irrelevant in the small $\rho$ limit, and hence
\beq  F = N \mu \log (1/\alpha) \left\{ \frac{(25-c)\mu^2}{96}
- 1 -\frac{4(2-\mu)}{\mu} \right\} + O(\log(N)) \label{freeN} \eeq
A similar calculation holds for $\rho=-2+\epsilon$; yet again the 
free energy takes the form of eq. (\ref{freeN}). When $c>1$, spike
formation is still favoured for small $\eta$.

\section{Conclusions and Discussion}

In this paper we discussed the stability of a random surface with
interacting spikes, and found that for a particular (unclumped) phase,
the transition to a spike dominated phase occurs as soon as $c$
becomes greater than unity. Together with the matrix-model
computations of David \cite{scenario} this provides strong evidence
for a sudden transition, devoid of an interpolating region.

A full proof that the $c>1$ region is dominated by
branched polymers is still lacking. While a simple-minded extension of
the above calculations suggests that the general structure of the free
energy remains the same as in (\ref{free4}) and (\ref{freeN}) (because
the Gamma function goes like $\Gamma(-n+\epsilon) \sim \epsilon^{-1}$,
this is especially apparent close to singularities of (\ref{bingo}))
we must be careful about such extrapolations, given that for $\rho<-2$
the spikes are no longer separated, but form clumps, casting doubts on
the spike-gas picture.

\vspace{1 truecm}
\noindent I thank W. E. Brown and J. F. Wheater for interesting
conversations on this topic. This work was partly supported by a grant
from {\sc PRAXIS XXI}.

\end{document}